\documentclass[aps,prl,amsmath,twocolumn,amssymb,footinbib,superscriptaddress]{revtex4}

\usepackage{amssymb,amsfonts,amsmath}

\usepackage{color}

\usepackage[apple mac]{inputenc}
\usepackage[english]{babel}
\usepackage{times}
\usepackage{latexsym}
\usepackage{fancyhdr}
\usepackage{verbatim}
\usepackage{graphicx}
\usepackage{epsf}
\usepackage{subfigure}
\usepackage{bm}
\usepackage{epstopdf}

\definecolor{Nathanblue}{rgb}{0.,0.24,0.51}
\newcommand{\blue}{\color{Nathanblue}}

\def\be{\begin{equation}}
\def\ee{\end{equation}}

\def\bs#1{\boldsymbol{#1}}

\begin{document}

\title{{\blue Extracting the Chern number from the dynamics of a Fermi gas: \\ Implementing a quantum Hall bar for cold atoms}}

\author{Alexandre Dauphin}
\email{adauphin@ulb.ac.be}
\affiliation{Center for Nonlinear Phenomena and Complex Systems - Universit\'e Libre de Bruxelles , 231, Campus Plaine, B-1050 Brussels, Belgium}\affiliation{Departamento de F\'isica Te\'orica I, Universidad Complutense, 28040 Madrid, Spain}

\author{Nathan Goldman}
\email{ngoldman@ulb.ac.be}
\affiliation{Laboratoire Kastler Brossel, CNRS, UPMC, ENS, 24 rue Lhomond, F-75005 Paris, France}

\begin{abstract}
We propose a scheme to measure the quantized Hall conductivity of an ultracold Fermi gas initially prepared in a topological (Chern) insulating phase, and driven by a constant force. We show that the time evolution of the center of mass, after releasing the cloud, provides a direct and clear signature of the topologically invariant Chern number. We discuss the validity of this scheme, highlighting the importance of driving the system with a sufficiently strong force to displace  the cloud over measurable distances while avoiding band-mixing effects. The unusual shapes of the driven atomic cloud are qualitatively discussed in terms of a semi-classical approach.
\end{abstract}

\date{\today}

\maketitle

\begin{figure}
\begin{center}
\includegraphics{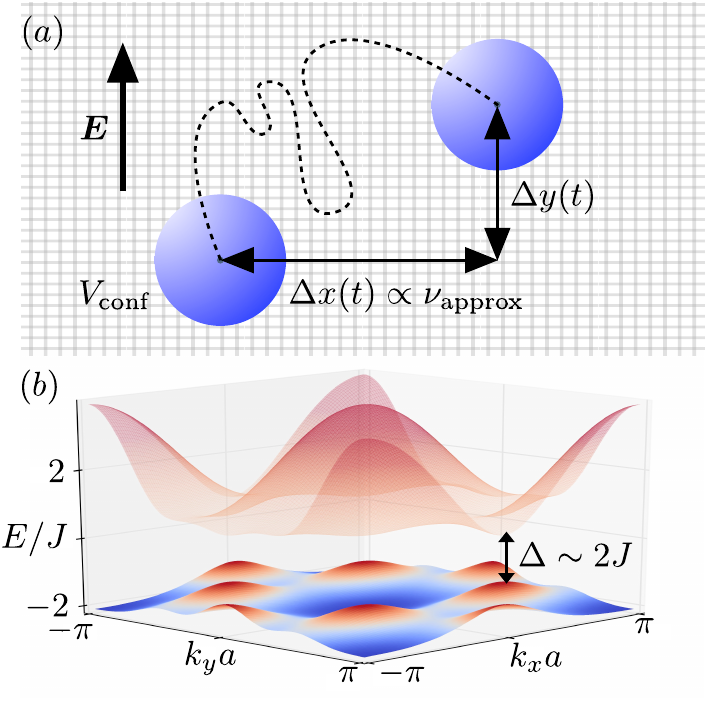}
\caption{(a) Atomic Chern insulating phase in the presence of an external force $\bs E =
E_y \bs 1_y$: the Chern number $\nu$ can be directly deduced from the transverse drift,
$\Delta x (t) \propto \nu_{\text{approx}} t$, which can be observed after releasing the
confining potential $V_{\text{conf}}$. (b) Band structure $E_{\pm} (\bs k)$ in the case
$J_2=0.3J$ and $\bs p=(0,4\pi/3a)$. The bulk energy gap is $\Delta=2J$; for all
configurations presented here $\Delta \sim 2J$ .}
\label{FIG1}
\end{center}
\end{figure}

The manifestation of topology in physical systems is no longer restricted to the realm of
solid-state setups, where the quantum Hall (QH) phases and topological insulators were
initially observed and created  \cite{Hasan2010,Qi2011}. Indeed, the ingredients
responsible for these topological phases, e.g. large magnetic fields or spin-orbit
couplings, have recently been engineered in cold-atom setups
\cite{Dalibard2011,Goldman:2013review,Lin2009b,Lin2011,Aidelsburger:2011,Cheuk2012,Wang2012,Aidelsburger:2013,Ketterle:2013,Ketterle:2013bis}. With such quantum simulators of
topological matter,  topological phases can be explored from a different perspective,
based on the unique probing and addressing techniques proper to cold-atom setups
\cite{Bloch:2012review,Maciejbook,Weitenberg}. \\
Topological phases are characterized by two fundamental properties
\cite{Hasan2010,Qi2011}: (a) a topological invariant $\nu$ associated with a bulk gap,
which is constant as long as the gap remains open, and (b) robust edge states whose
energies are located within the bulk gap. A first manifestation of topology was discovered
in the QH effect \cite{Thouless1982,Kohmoto:1985}, where the Hall conductivity is exactly
equal to the topological Chern number $\nu \! \in \! \mathbb{Z}$ in units of the
conductivity quantum $\sigma_0$, i.e. $\sigma_H \! = \! \nu \sigma_0$. In solid  materials
subjected to large magnetic fields, the quantized Hall conductivity is measured through
the transport equation $\bs{j} \! = \! \sigma \bs{E}$, where $\bs E \! = \! E_y \bs 1_y$
is an electric field and where a non-zero transverse conductivity $\sigma_{H} \! = \!
\sigma_{xy}$ signals the Hall current $j_x$ generated by the magnetic field
\cite{Klitzing1986}. Engineering the analogue of a QH experiment with cold atoms subjected
to synthetic magnetic fields \cite{Dalibard2011,Goldman:2013review} would require to drive the system along a
given direction, and to measure the Hall current in the transverse direction
\cite{Goldman:2007}. For the analogy to be complete, reservoirs should be connected to the
cold-atom systems, in order to inject and retrieve the driven particles. Although
mesoscopic conduction properties have been demonstrated in an ultracold Fermi gas
``connected" to two reservoirs \cite{Brantut2012}, such a scheme would add a considerable
complexity to the demanding setup that generates the synthetic magnetic field. To overcome
this issue, strategies have been proposed to evaluate the topological invariant $\nu$ by
other means, based on hybrid time-of-flight \cite{Wang:2013}, Bloch oscillations
\cite{Price:2012,Liu:2013}, the Zak's phase measurement \cite{Atala:2012,Abanin:2012}, and
density imaging \cite{Alba:2011,Goldman:2013njp,Umucallar:2008,Shao:2008,Zhao:2011}.
Modulations of the external confining potential has already revealed Hall-like behaviors
in the presence of  synthetic magnetic fields \cite{LeBlanc2012,Pino:2013}. Topological
edge states that are expected to be present when $\nu \! \ne \! 0$ could also be
visualized
\cite{GoldmanDalibard:2012,Goldman:2012prl,Killi:2012,Scarola:2007,Liu:2010,Stanescu2010}.
\\
In this Letter, we introduce a scheme to directly measure the Hall conductivity of a Fermi
gas trapped in a 2D optical lattice and driven by a constant external force $\bs{E} \! =
\! E_y \bs{1}_y$ (e.g. a lattice acceleration \cite{Jaksch:2003}). Our method is based on
the possibility to prepare the atomic gas in a Chern insulating phase, and to image the
time evolution of its center-of-mass (CM) $\bs x (t)$ after suddenly releasing the
confining potential [Fig. \ref{FIG1} (a)]. This  cold-atom QH measurement leads to a
satisfactory measure of the Chern number $\nu$ under two conditions: (1) the force $E_y$
should be strong enough to generate a measurable displacement after a realistic
experimental time, e.g. $\bs x (t) \! \sim \! 10 a$, where $a$ is the lattice spacing; (2)
the force should be small compared to the topological bulk gap $E_y a \! \ll \! \Delta$,
to avoid band-mixing processes \cite{Zener}. Under those assumptions, our method is valid
for \emph{any} cold-atom system hosting QH
\cite{Shao:2008,Stanescu:2009,Stanescu2010,Liu:2010,Goldman:2013Haldane,Hafezi:2007} or
quantum spin Hall phases \cite{Goldman:2010,Hauke:2012}.\\
\label{sect:model}
To illustrate the method, we consider a non-interacting Fermi gas in a 2D brick-wall
optical lattice \cite{Tarruell2011} with complex nearest-neighbor (NN) hopping $J e^{i
\theta}$, and real next-nearest-neighbor (NNN) hopping $J_2 < J$. This system can be
realized through shaking techniques \cite{Hauke:2012}, or by trapping atoms in two
internal states, using state-dependent optical lattices, and by inducing the NN hopping
through laser-coupling
\cite{Jaksch:2003,Aidelsburger:2011,Gerbier:2010,Alba:2011,Goldman:2013njp}. The
second-quantized Hamiltonian is taken to be \cite{Alba:2011,Goldman:2013njp}
\be
\hat H=-J \sum_{\langle i,j \rangle} e^{i \bs p \cdot (\bs r_i + \bs r_j)/2} \hat
c_{i}^{\dagger} \hat c_j - J_2 \sum_{\langle \langle k,l \rangle \rangle}  \hat
c_{k}^{\dagger} \hat c_l,
\ee 
where $\bs p$ is the recoil momentum associated with the laser coupling
\cite{Jaksch:2003,Gerbier:2010}, and where $\hat c_{k}^{\dagger}$ creates an atom at
lattice site $\bs r_k$. This system realizes the two-band Haldane model
\cite{Haldane:1988}: for certain values of $\bs p$ and $J_2/J$, a topological bulk gap
opens with Chern number $\nu=\pm 1$ \cite{Alba:2011,Goldman:2013njp}, leading to anomalous
QH phases
 \cite{Haldane:1988}. The bulk energy gap can be as large as $\Delta = 2 J$ [Fig.
\ref{FIG1} (b)]. Below, we describe a scheme to measure the Chern number $\nu = \pm 1$,
assuming that the atomic gas can be prepared in such a phase \cite{note_prep}. 

Let us first assume that a single atom is confined in an optical lattice of size $L \times
L$, where $L$ is the number of unit cells along each direction. In the presence of a force
directed along $y$, $\hat H_{\text{electric}}=  - E_y \sum_j y_j \hat c_{j}^{\dagger} \hat
c_j$, the velocity in a state $\vert u_{(-)} , \bs k \rangle$ of the lowest band $E_{-}
(\bs k)$ with quasi-momentum $\bs k$ is  \cite{Xiao2010}
\begin{align}
& v_x(\bs k)  = v_{\text{band}}^x + v_{\mathcal{F}}^x =  \frac{\partial E_{-} (\bs
k)}{\hbar \partial k_x} - \frac{i}{\hbar} E_y \, \mathcal{F}_{xy}^{(-)} (\bs k),  \notag
\\
& v_y(\bs k)  =v_{\text{band}}^y= \frac{\partial E_{-} (\bs k)}{\hbar \partial k_y}
,\label{velocity}
\end{align}
where $\mathcal{F}_{xy}^{(-)} (\bs k)= \langle \partial_{k_y} u \vert \partial_{k_x} u
\rangle - \langle \partial_{k_x} u \vert \partial_{k_y} u \rangle$ is the Berry's
curvature associated with the state. Completely filling the lowest band $E_{-} (\bs k)$
and taking the limit $L \rightarrow \infty$ yields the relation for the current density in
the QH phase
\be
 j_x= - ( \nu / h) \, E_y \, , \quad   j_y   =0, \label{chern_current}
\ee 
where $\nu= (i/2 \pi) \int_{\text{BZ}} \mathcal{F}_{xy}^{(-)} \text{d} \bs k$ is the Chern
number of the lowest band \cite{Kohmoto:1985}, and where BZ denotes the first Brillouin
zone. In order to avoid measuring the current, which would require connecting reservoirs
to the system, we follow an alternative strategy. We initially confine the system in a
region of size $L_0<L$ using a confining potential, $\hat H_0=\hat H + \hat V_{\text{conf}}$, and we set the Fermi
energy $E_{\text{F}}$ inside the topological bulk gap, hence filling the lowest band
$E_{-} (\bs k)$ completely. At time $t=0$, we suddenly remove the potential
$V_{\text{conf}}$ and add the force. After the quench, all the initial states project onto
the eigenstates of the final Hamiltonian $\hat H_{\text{tot}}=\hat H_0 + \hat
H_{\text{electric}} -  \hat V_{\text{conf}}$, uniformly populating the lowest band $E_{-}
(\bs k)$. Edge states lying within the bulk gap partially project unto states of the
highest band $E_{+} (\bs k)$, but this effect is negligible  \cite{sup_mat}. Taking into
account the velocity \eqref{velocity} associated with all the occupied states, and
neglecting any contribution from the highest band, we find that the CM follows the
equations of motion
\be
x (t)= - (a^2 t E_y/ \pi \hbar ) \, \nu_{\text{approx}} \, , \quad
y(t)=0,\label{chern_position}
\ee
where $\nu_{\text{approx}}$ is a discretized expression for the Chern number that
converges towards $\nu$ as $L_0 \rightarrow \infty$ \cite{notediscr}, and where we used
the fact that each unit cell has an area $2 a^2$. 
Importantly, the initial filling of the lowest band cancels the undesired contribution of
the band velocity $\bs v_{\text{band}} $. This constitutes a significant advantage with
respect to proposals based on bosonic wave packets, where this effect must be annihilated
by other means to measure $\nu$ \cite{Price:2012}.

\begin{figure}
\begin{center}
\includegraphics{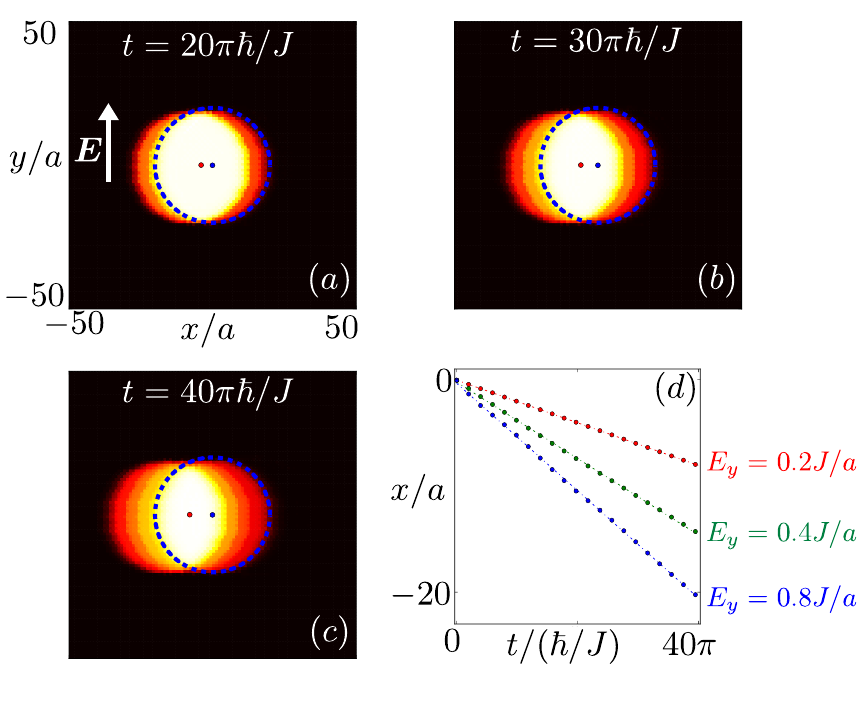}
\caption{Time-evolving particle density $\rho (\bs x,t)$ of a QH atomic cloud, for
$J_2\!=\!0.3J$ and $\bs p\!=\!(0,4\pi/3a)$. (a)-(c) The system is initially confined in a
disk of radius $r_0\!=\!20a$, and is driven by a constant force $\boldsymbol{E}\!=\!0.2
J/a \boldsymbol{1}_y$. The blue dashed circles and dots denote the initial distribution of
the cloud and its CM $\bs x(t\!=\!0)$. The red points show the CM $\bs x(t)$ at time $t
>0$. The figure (d) shows the displacement $x(t)$ for increasing values of the driving
force $\boldsymbol{E}\!=\!E_y \bs{1}_y$. Applying Eq. \eqref{chern_position} to these curves yields
$\nu_\text{approx}\!=\!1.00$ ($E_y\!=\!0.2J/a$), $\nu_\text{approx}\!=\!0.90$
($E_y\!=\!0.4J/a$) and $\nu_\text{approx}\!=\!0.63$ ($E_y\!=\!0.8J/a$): the Chern-number measurement
breaks down above the critical force $E_y^{\text{c}} \!\simeq\! 0.3 J/a$ for $\Delta =
2J$.}
\label{FIG2}
\end{center}
\end{figure}

We now simulate such a protocol and discuss the regimes in which the measured quantity
$\nu_{\text{approx}}$ provides a satisfactory evaluation of the Chern number $\nu$,
revealing an unambiguous signature of topological order.  In order to minimize 
band-mixing effects  \cite{Price:2012}, we set the model parameters to the values
$J_2=0.3J$ and $\bs p=(0,4\pi/3a)$ that maximize the spectral gap $\Delta = 2 J$ [Fig.
\ref{FIG1} (b)]. We initially confine the system with a perfectly sharp circular potential
$V_{\text{conf}} (r) = J (r/r_0)^{\infty}$, which can now be created in experiments
\cite{Meyrath:2005,Gaunt:2012,Blochdiscussion}; smooth confinements are discussed in
\cite{sup_mat}. In this configuration, we set the Fermi energy $E_{\text{F}} \gtrsim
\text{max} (E_{-} (\bs k))$ so as to fill the lowest band while limiting the population of
edge states \cite{sup_mat}. At time $t=0$, we suddenly remove the confinement
$V_{\text{conf}}$ and act on the system with a weak force $\bs E = 0.2 J/a \bs{1}_y$.
Figures \ref{FIG2} (a)-(c) show the time evolution of the particle density $\rho (\bs
x,t)$, demonstrating a clear drift of the cloud along the transverse direction $x$; 
after a typical time $t^* = 40 \pi \hbar/J$ \cite{note:time}, this CM displacement is
$\vert \bs x (t^*) \vert = 8 a \bs 1_x$, which is detectable using available
high-resolution microscopy \cite{Brantut2012,Weitenberg}. Figure \ref{FIG2} (d) shows the
displacement $x(t)$ as a function of time, for different values of the force. For $E_y =
0.2 J/a$, the system is driven in the linear-response regime, and the CM follows the
constant motion  \eqref{chern_position}. A linear regression applied to the data $x(t)$
yields a precise value for the measured Chern number $\nu_{\text{approx}} = 1.00$.
Increasing the force allows to enlarge the displacement, which is desirable to improve the
detection; however, it is also crucial to avoid non-linear effects in order to measure
$\nu$ adequately through Eq. \eqref{chern_position}. For $E_y=0.4 J/a$, the  displacement
at time $t^*$ is $\vert \bs x (t^*) \vert = 14 a \bs 1_x$, but the measured quantity
$\nu_{\text{approx}} = 0.90$ already signals the breakdown of the single-populated-band
approximation; the transfer to the higher band could be confirmed through band-mapping
techniques \cite{Tarruell2011}. For a significantly larger force $E_y=0.8 J/a$, a clear
Hall drift is still observed, however, the measured quantity $\nu_{\text{approx}} \!=\!
0.63$ largely deviates from the quantized value. From Fig. \ref{FIG2}, we conclude that a
moderate force $E_y \approx \pm 0.3 J/a$ constitutes a good compromise, allowing one to
measure a robust Chern number $\nu_{\text{approx}} \!=\! \pm 1.0$ through a CM
displacement of a few tens of lattice sites \cite{single-site}, under realistic times
$t\!\sim\!10\!-\!100 \text{ms}$ \cite{note:time}. In this weak-force regime, the
measurement is robust against  perturbations preserving the band structure, in agreement
with the topological nature of the Chern number. The flatness of the lowest band in Fig.
\ref{FIG1} (b) does not influence our result, as the displacement $\bs x(t)$ only relies
on $\nu$, and not on the band velocity $\bs v_{\text{band}} =(1/\hbar) \partial_{\bs k}
E_-(\bs k)$. 

\begin{figure}
\begin{center}
\includegraphics{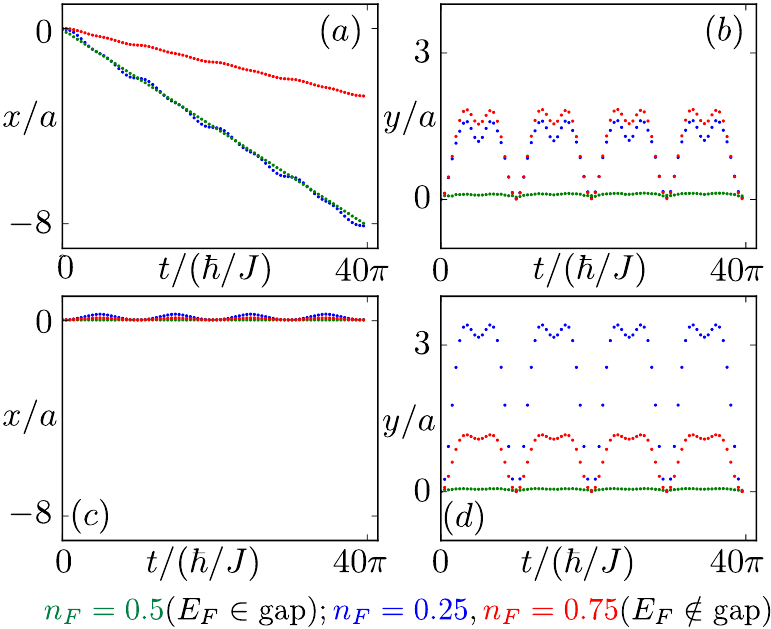}
\caption{Motion of the CM $\bs x(t)$ for different filling factors $n_F$. (a)-(b) The
topological case ($J_2=0.3J,\lambda_\text{stag}=0$); (c)-(d) the trivial case
($J_2=0,\lambda_\text{stag}=1.0J$). The force is $E_y= 0.2 J/a$ and the size of the bulk
gap $\Delta=2J$ is the same for both cases. }
\label{FIG3}
\end{center}
\end{figure}
\begin{figure}
\begin{center}
{\scalebox{0.9}
{\includegraphics{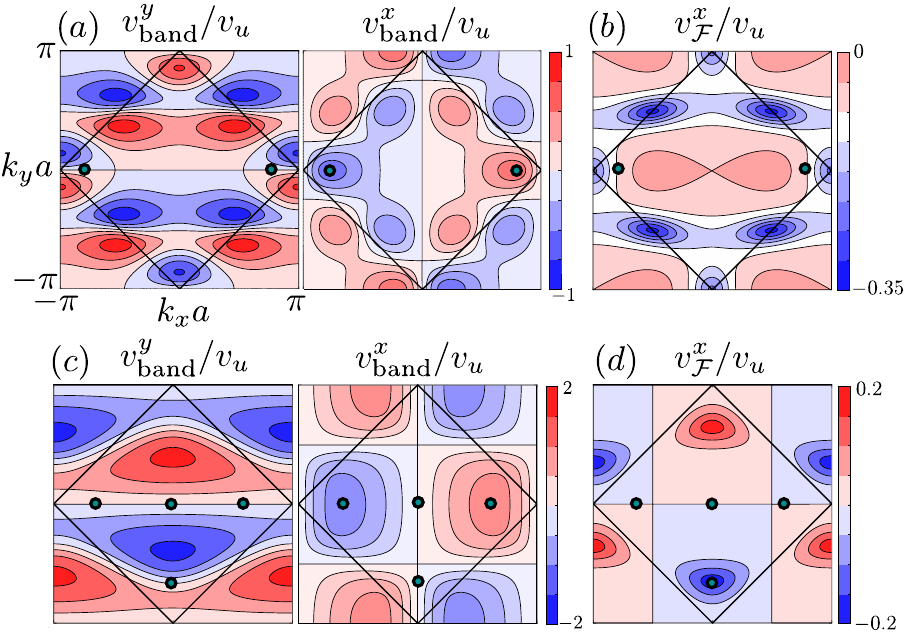}}}
\caption{The band and Berry's velocities in units of $v_u=Ja/\hbar$ for  $E_y=0.2 J/a$.
(a)-(b) The topological case ($J_2=0.3J,\lambda_\text{stag}=0$); (c)-(d) the trivial case
($J_2=0,\lambda_\text{stag}=1.0J$). 
to describe the dynamics of the atomic cloud in Fig. \ref{FIG5}.
}
\label{FIG4}
\end{center}
\end{figure}

We now study the stability of our method against variations of the atomic filling factor.
This effect is investigated in Figs. \ref{FIG3} (a)-(b), which compare the time evolution
of the CM for different fillings $n_F=1/4,1/2, 3/4$. The half-filling case $n_F=1/2$ (i.e.
the QH phase $E_{\text{F}} \in \Delta$) shows the constant Hall drift $x(t)$ dictated by
Eq. \eqref{chern_position} and the immobility along the driven direction $y(t)\approx 0$.
When $n_F=1/4$, the lowest band is partially filled and the system behaves as a metal: a
clear motion along $y$ accompanied with Bloch oscillations is observed [Fig.  \ref{FIG3}
(b)]. Interestingly, the motion along the transverse direction $x$ is characterized by an
almost constant velocity, which when fitted with the filled-band expression
\eqref{chern_position} yields an approximatively quantized value  $\nu_{\text{approx}} =
1.03$; this results from the fact that the evolving occupied states contribute
significantly to the total Berry's velocity $\sum_{\bs k} v^x_{\mathcal{F}}(\bs k) \propto
\nu$, while their contribution to the band velocity $\sum_{\bs k} v^x_{\text{band}} \simeq
0$ vanishes by symmetry  
[Fig. \ref{FIG4}]. When $n_F=3/4$, the upper band $E_{+}(\bs k)$ is partially filled and
the system also behaves as a metal along the $y$ direction. Here, the contribution of
high-energy states strongly affects the motion along the $x$ direction, which when fitted
with Eq. \eqref{chern_position} yields a non-quantized value $\nu_{\text{approx}} \approx
0.3$: the contribution of the Berry's curvature $\mathcal{F}^{(+)}_{xy} = -
\mathcal{F}^{(-)}_{xy} $ associated with $E_{+}(\bs k)$ 
 spoils the evaluation of the Chern number. 
We have verified that the measured quantity $\nu_{\text{approx}} \simeq \nu$ 
remains robust for small filling variations around the QH phase, $n_F \approx 1/2$
\cite{sup_mat}.\\
\label{sect:trivial}
The Chern number characterizes the topological class of the system
\cite{Hasan2010,Qi2011,Kohmoto:1985}, and thus, it distinguishes between a trivial
insulating phase ($\nu=0$) and a topological insulating phase ($\nu \ne 0$). To further
evaluate the efficiency of our method, we compare the CM motion discussed above with a
system configuration  corresponding to a trivial topological order. To do so, we introduce
a staggered potential $\hat H_{\text{stag}}=  \lambda_{\text{stag}} \sum_j (-1)^{j}
\hat c_{j}^{\dagger} \hat c_j$, which adds an onsite energy $\pm \lambda_{\text{stag}}$
alternatively along both spatial directions. This perturbation opens a trivial bulk gap
 with $\nu=0$ \cite{Haldane:1988,Alba:2011,Goldman:2013njp}. We show in Figs. \ref{FIG3}
(c)-(d) the CM motion for this trivial configuration ($J_2=0$, $\lambda_{\text{stag}}=J$),
considering different filling factors; $\lambda_\text{stag}$ is chosen such that the width
of the bulk gap $\Delta = 2 J$ is the same as for the topological case [Figs. \ref{FIG3}
(a)-(b)]. At half filling, the system is immune to the external force, $\bs x (t) \approx
0$, in agreement with the behavior of an insulator; we find $\nu_{\text{approx}}=0$ with
less than $1\%$ error. In the metallic phases, the cloud performs Bloch oscillations along
the driving direction $y$, while no Hall transport is observed, $x (t) \simeq 0$; see also
\cite{sup_mat}. 

\label{sect:classic}

\begin{figure}
\begin{center}
{\scalebox{0.9}
{\includegraphics{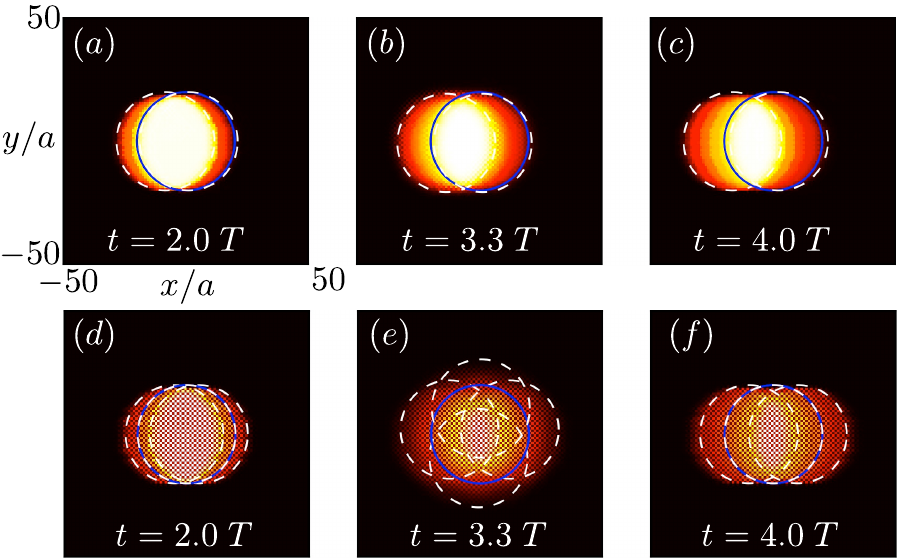}}}
\caption{Time-evolving spatial density $\rho (\bs x,t)$ for $E_y=0.2 J/a$ and
$E_{\text{F}} \in \Delta$. (a)-(c) The topological phase
($J_2=0.3J,\lambda_\text{stag}=0$); (d)-(f) the trivial insulating phase
($J_2=0,\lambda_\text{stag}=1.0J$). The plain circle represents the initial condition. The
dashed circles represent the semi-classical trajectories of the wave-packets with initial
momenta $\bs k^0$ shown in Fig. \ref{FIG4}. Time is expressed in units of the period
$T=10\pi \hbar/J$. See also \cite{sup_mat}.}
\label{FIG5}
\end{center}
\end{figure}

The dynamics of the QH atomic gas is characterized by two different effects: (a) the CM
displacement captured by Eq. \eqref{chern_position} as discussed above; and (b) the
dynamical deformations of the cloud. The latter effect arises as an interplay between the
band structure and the force applied to the system. These deformations can be
qualitatively described through a semi-classical picture \cite{Xiao2010,Price:2012}, in
which the dynamics of the cloud is decomposed into  wave packets $\psi (\bs x(t) , \bs
k_n(t))$, localized around the CM $\bs x (t)$ and the many quasi-momenta $\bs k_n(t) \in
\text{BZ}$. When a force is applied along the $y$ direction, the quasi-momentum of a
single wave packet initially localized around $\bs k^0$ evolves according to $k_x \! = \!
k_x^0$ and $k_y \! = \! k_{y}^0+E_y t/\hbar $. The real space evolution of each wave
packet is dictated by $\dot{\bs x} (t) \!  = \!  \bs v (\bs k)$, where the velocity $\bs v
\!  = \!  \bs v_{\text{band}} \!  + \!  \bs v_{\mathcal{F}}$ is given in Eq.
\ref{velocity} [see Fig. \ref{FIG4}]. We now discuss the deformations of the cloud by
solving these semi-classical equations independently, for a few chosen values of $\bs k^0$
that capture the essential diffusion effects. The motion along the $y$ direction is
entirely determined by the band velocity $ v_{\text{band}}^y (\bs k)$ [Figs. \ref{FIG4}
(a),(c)], which leads to Bloch oscillations: after a full period $T \!  = \!  2 \pi \hbar/
a E_y$, all the wave packets return to their initial position $y(T) \!  = \!  0$ and $\bs
k(T) \! = \!  \bs k^0$ [Fig. \ref{FIG5}]. The motion taking place along the transverse
direction $x$ is more exotic, as it is influenced by the Berry's velocity $\bs
v_{\mathcal{F}}$ \cite{Price:2012}. First, the motion of a wave packet initially at $\bs
k^0 \!  = \!  0$ is characterized by a finite Berry's velocity and a zero band velocity
$v_{\text{band}}^x \!  = \!  0$ [Fig. \ref{FIG4}]. In the topological case $\nu \!  = \! 
1$, the Berry's velocity is always negative, which leads to a net drift along $x$. This
analysis can be readily extended to wave packets initially centered around other  momenta
$\bs k^0 \!  \ne \!  0 \!  \in \!  \text{BZ} $, whose transverse drift are affected by the
band velocity $v_{\text{band}}^x \!  \ne \!  0$. By symmetry, the initial conditions shown
in Fig. \ref{FIG4}, corresponding to $k^0_x \!  < \!  0$ and $k^0_x \!  > \!  0$,  evolve
with the same Berry's velocity but opposite band velocity; these two wave packets undergo
a net drift along \emph{opposite} directions after each period $T$ [Figs. \ref{FIG4} \!-\!
\ref{FIG5}]. These typical opposite drifts yield the progressive broadening of the cloud
along the $x$ direction. Combining this net diffusion together with the Bloch oscillations
leads to unusual shapes of the cloud at arbitrary times $t \ne T \times \text{integer}$
\cite{sup_mat}.

We emphasize that our Chern-number measurement does not rely on the methods used to
generate the topological band structure; thus, it can be applied to any cold-atom setup
characterized by non-trivial Chern numbers. In particular, this scheme could be applied to
distinguish between different Chern insulators with $\vert \nu \vert \ge 1 \in
\mathbb{N}$. Moreover, our method could be extended to the case of $Z_2$ topological
phases \cite{Goldman:2010,Hauke:2012,interactions}, where the spin Chern number could be
deduced by subtracting the CM displacements associated with the two spin species,
$\nu_{\text{spin}} \propto \bs x_{\uparrow} - \bs x_{\downarrow}$. Finally, our scheme
could be applied to QH photonic systems \cite{photonic}.


%
%
%
%
%

We thank the FRS-FNRS (Belgium) and the ULB for financial support, and M. A. Martin Delgado, P.
Debuyl, P. Gaspard, M. M\"uller, L. Tarruell, F. Gerbier, J. Dalibard, J. Beugnon, I.
Bloch, D. Greif, and H. Price for stimulating discussions and support.

\vspace{-0.5cm}

\bibliographystyle{apsrev}

\begin{thebibliography}{40}
\bibitem{single-site} We note that single-site resolution microscopy \cite{Weitenberg}
will generally be required to measure an ``integral" value $\nu_{\text{approx}}=\nu +
\delta$, with $\nu \in \mathbb{Z}$ and $\delta < 0.01$.
\end{thebibliography}


\newpage

\onecolumngrid
\appendix

\begin{center}
{\large \bf{{\blue Supplementary Material}}}\\
\end{center}
\vspace{0.5cm}

\begin{flushleft}
{\bf {\blue Appendix A}}: Effects of the populated edge states on the dynamics \\
\vspace{0.25cm}
{\bf {\blue Appendix B}}: Bloch oscillations and dynamics at arbitrary times: flat bands vs dispersive bands \\
\vspace{0.25cm}

{\bf {\blue Appendix C}}: Competition between trivial and non-trivial topological phases\\
\vspace{0.25cm}
{\bf {\blue Appendix D}}: The Chern number measurement using smooth confinements\\
\vspace{0.25cm}

{\bf {\blue Appendix E}}: Releasing the confinement along the transverse direction only
\end{flushleft}

 \section{\label{sec:edgeeffects} Appendix A: Effects of the populated edge states on the dynamics}

\begin{figure}[h!]
\begin{center}
\includegraphics{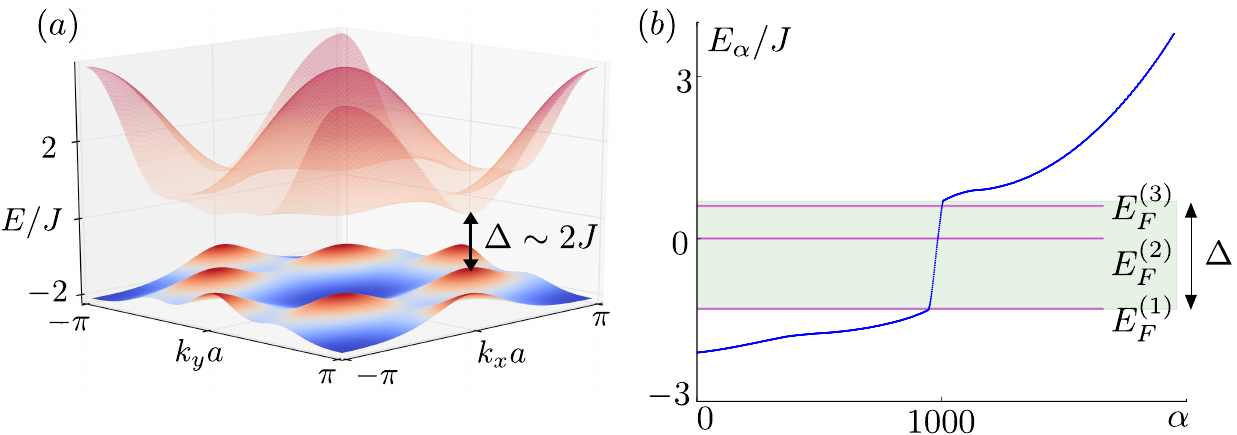}
\caption{(a) Bulk band structure $E_{\pm} (\bs k)$ in the case $J_2=0.3J$ and $\bs p=(0,4\pi/3a)$, see main text. The bulk energy gap is $\Delta=2J$; for all configurations presented in this Appendix $\Delta = 2J$. (b) Discrete energy spectrum for the same system configuration, in the presence of an infinitely abrupt circular potential $V_{\text{conf}} (r)$ with radius $r_0=25a$. The three Fermi energies $E_F^{(1,2,3)}$ are considered in this Appendix in order to study the effects of the edge states on the dynamics. The label $\alpha$ classifies the energies in increasing order, $E_{\alpha} < E_{\alpha+1}$.}
\label{fig:seca_fig1}
\end{center}
\end{figure}

In the main text, we considered a Fermi gas initially confined by an infinitely abrupt potential $V_{\text{conf}}(r)=(r/r_0)^{\infty}$  and prepared in a quantum Hall (QH) phase. The cold-atom system is characterized by the band structure illustrated in Fig. \ref{fig:seca_fig1} (a) and the Fermi energy is set within the bulk gap denoted $\Delta$. In this configuration, the lowest band $E_{-} (\bs k)$ with Chern number $\nu=1$ is totally filled. At time $t=0$, the confinement is suddenly released $V_{\text{conf}}(r) = 0$ and a force is added along the $y$ direction, $\bs E = E_y \bs 1_y$. 
Assuming that all the initially populated states project unto states of the lowest energy band $E_{-} (\bs k)$, which is a valid hypothesis as far as the bulk states are concerned (see below), we obtained the equations of motion for the center of mass,
\be
x (t)= - (a^2 t E_y/ \pi \hbar ) \, \nu_{\text{approx}} \, , \quad y(t)=0, \notag
\ee
where $\nu_{\text{approx}} \approx \nu$, see main text. Clearly, these equations neglect the fact that edge states, whose energies are located within the bulk gap, are initially populated. Indeed, the edge states that are spatially localized in the vicinity of the confining radius $r \approx r_0$, will potentially project on the (many) bulk states associated with the two bulk bands $E_{\pm} (\bs k)$. Since the Chern numbers of the two bands are opposite $\nu^{(+)} = - \nu^{(-)}$, the contribution of the initially populated edge states to the dynamics can potentially perturb the Chern number measurement described in the main text. It is the scope of this Appendix to show to what extend their contribution can indeed be neglected. \\

The two-band spectrum $E_{\pm} (\bs k)$ shown in Fig. \ref{fig:seca_fig1} (a), which has been obtained by considering periodic boundary conditions, does not take into account the edge states that are present in the experimental setup: the non-zero Chern number $\nu =  1$ guarantees the presence of edge states that are spatially localized at $r=r_0$, and whose energies are located within the bulk gap. These edge states are visible in the spectrum $E_{\alpha}$ represented in Fig. \ref{fig:seca_fig1} (b), which has been obtained for a trapped system with circular confining potential $V_{\text{conf}}(r)=(r/r_0)^{\infty}$ and $r_0=25 a$. 

\begin{figure}[h!]
\begin{center}
\includegraphics{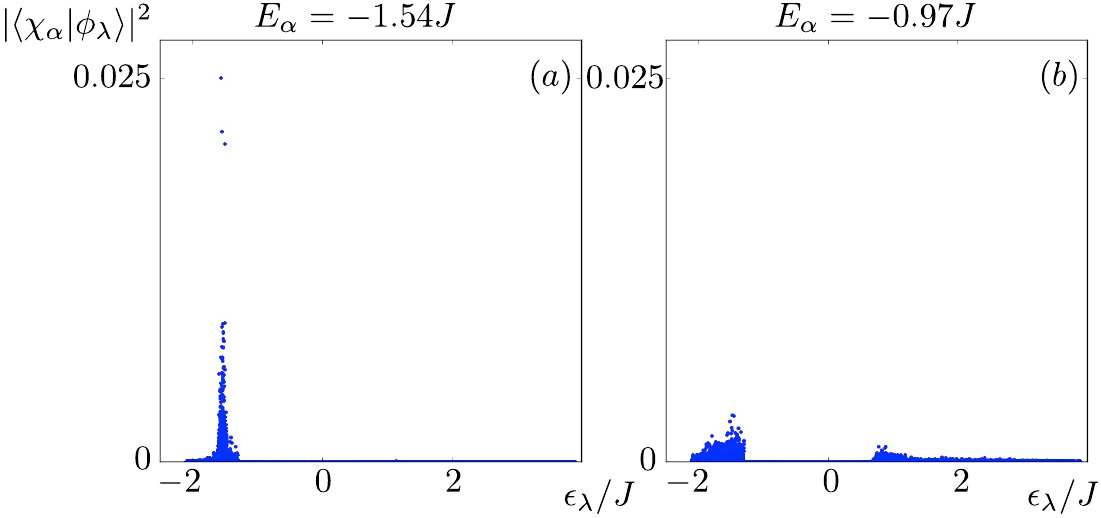}
\caption{(a) Projection of a bulk state $\vert \chi_{\alpha} \rangle$ of the confined system ($V_{\text{conf}} \ne 0$), with energy $E_\alpha=-1.54J$, onto the states $|\phi_\lambda \rangle$ of the unconfined system ($V_{\text{conf}} = 0$).  (b) Projection of an edge state $\vert \chi_{\alpha} \rangle$ of the confined system with energy $E_\alpha=-0.97J$: after releasing the trap, the edge states project onto bulk states associated with the two bulk bands $E_{\pm} (\bs k)$.   }
\label{fig:seca_fig3}
\end{center}
\end{figure}

We now consider three different values for the Fermi energy: $E_F^{(1)}$ is set right above the lowest bulk band $E_{-} (\bs k)$, $E_F^{(2)}=0$ is located well inside the bulk gap, and $E_F^{(3)}$ is set right below the highest band $E_{+} (\bs k)$. Note that the value $E_F^{(1)}=-1.3 J$ corresponds to the situation considered in the main text. When suddenly releasing the confinement $V_{\text{conf}}(r)$, a bulk state $\chi_{\alpha}$ with energy $E_{\alpha} \in E_{-} (\bs k)$ will project on bulk states $\phi_{\lambda}$ with energies $\epsilon_{\lambda} \approx E_{\alpha}$, as shown in Fig. \ref{fig:seca_fig3} (a). On the contrary, an edge state with energy $E_{\alpha} \in \Delta$ will project on bulk states with energies $\epsilon_{\lambda} \in E_{-} (\bs k)$ and also on bulk states with energies $\epsilon_{\lambda} \in E_{+} (\bs k)$, as shown in Fig. \ref{fig:seca_fig3} (b). As a corollary, the population of states lying in the highest band $E_{+} (\bs k)$ and taking part in the dynamics is reduced by setting the Fermi energy close to the band edge $E_F \approx E_F^{(1)}$, while this undesired population is increased for higher Fermi energies  $E_F= E_F^{(2,3)} > E_F^{(1)}$. 

\begin{figure}[h!]
\begin{center}
\includegraphics{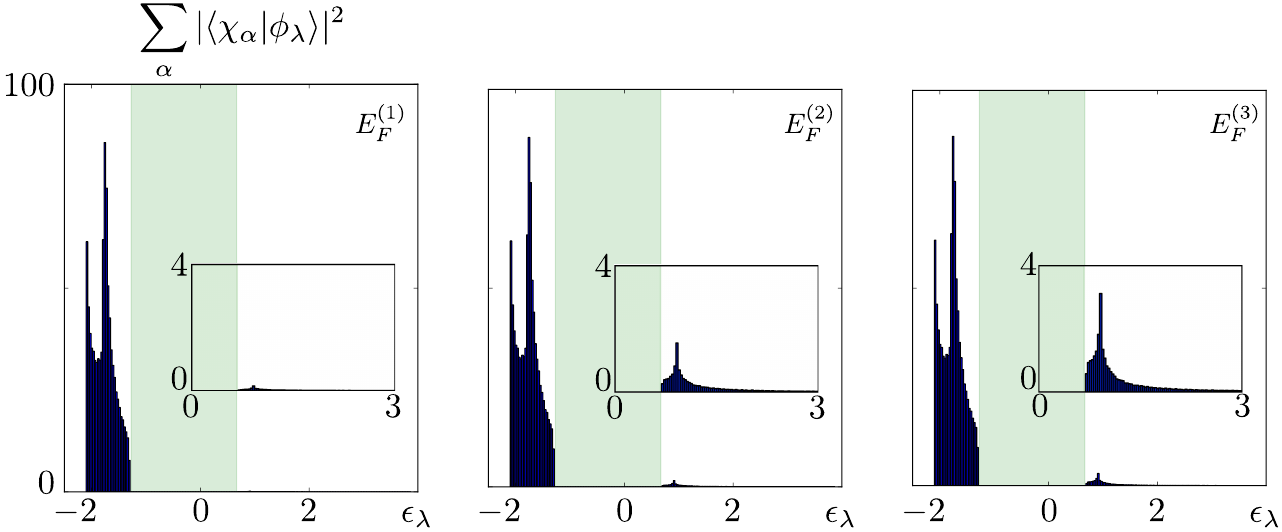}
\caption{Population of the states $\phi_{\lambda}$ with energy $\epsilon_{\lambda}$ after releasing the confinement $V_{\text{conf}}$, for the different values of the Fermi energy $E_F=E_F^{(1,2,3)}$ shown in Fig. \ref{fig:seca_fig1}. In all figures, the green shaded region corresponds to the energy bulk gap $\Delta$. The partial filling of the highest band $E_{+} (\bs k)$ is highlighted for all cases in small boxes.}
\label{fig:seca_fig2}
\end{center}
\end{figure}

The states populations $\mathcal{P}_{\lambda} = \sum_{E_{\alpha}<E_F} \vert \langle \chi_{\alpha} \vert \phi_{\lambda} \rangle \vert^2$ after the quench are represented in Fig. \ref{fig:seca_fig2} for $E_F= E_F^{(1,2,3)}$. Here, $\chi_{\alpha}$ [resp. $\phi_{\lambda}$] denotes the eigenstate with energy $E_{\alpha}$ [resp. $\epsilon_{\lambda}$] before [resp. after] the quench. From Fig. \ref{fig:seca_fig2}, we deduce that the population of the highest band $E_{+} (\bs k)$ is highly limited, even in the extreme case where all the edge states are initially filled, i.e. when $E_F= E_F^{(3)}$.

\begin{figure}[h!]
\begin{center}
\includegraphics{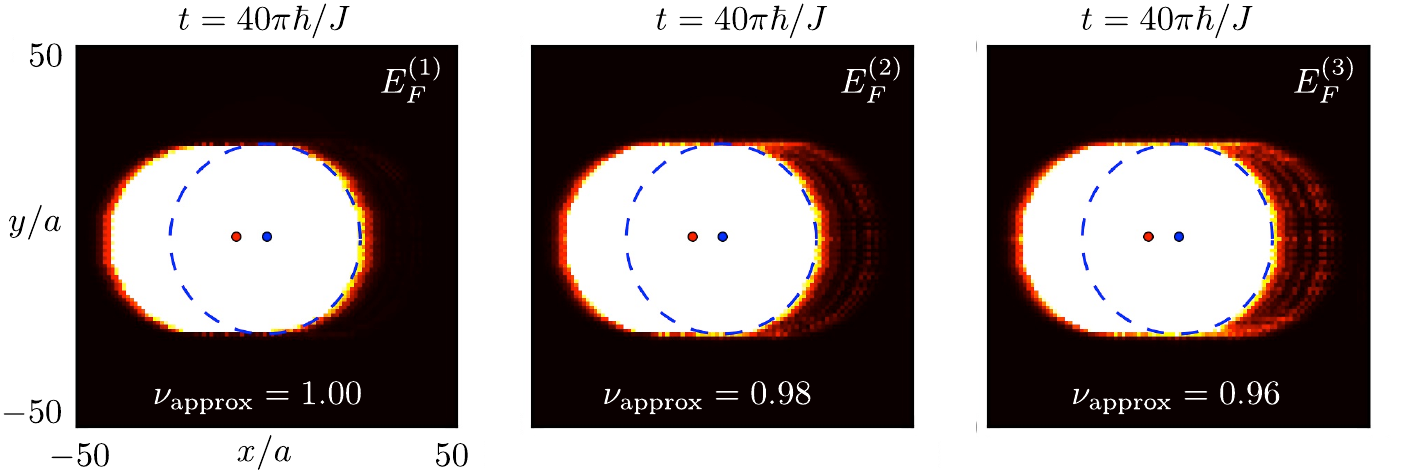}
\caption{Spatial density $\rho (\bs x,t)$ at time $t=40\pi \hbar/J$ for the three Fermi energies $E_F=E_F^{(1,2,3)}$ presented in Fig. \ref{fig:seca_fig1} (b). The dashed blue circle represents the atomic cloud at $t=0$, while the blue [resp. red] dots show the position of the center of mass at time $t=0$ [resp. $t=40\pi/\hbar$]. The non-zero projection onto the upper band $E_{+} (\bs k)$ gives rise to a small counter-propagating motion, due to the opposite value of the Berry's curvature $\mathcal{F}^{(+)}_{xy} = - \mathcal{F}^{(-)}_{xy}$, see also main text.}
\label{fig:seca_fig4}
\end{center}
\end{figure}

We now illustrate how the population of the highest band modifies the dynamics of the cloud, and thus how it affects the Chern number measurement. The time-evolving density is shown in Fig. \ref{fig:seca_fig4}, for the three different values of the Fermi energy $E_F= E_F^{(1,2,3)}$ discussed above. By increasing the contrast of the corresponding density plots, we observe the appearance of a few particles that move to the right, i.e. in the direction opposite to the overall Hall drift. These few states, whose population increases with the Fermi energy, are associated with the highest band $E_{+} (\bs k)$ and they have an opposite Berry's velocity $v_{\mathcal{F}}^{(+)} = - v_{\mathcal{F}}^{(-)}$. We find that these few counter-propagating states only slightly affect the center-of-mass displacement: the Chern numbers evaluated from the dynamics are $\nu_{\text{approx}} =1.00$ ($E_F= E_F^{(1)}$), $\nu_{\text{approx}} =0.98$ ($E_F= E_F^{(2)}$) and   $\nu_{\text{approx}} =0.96$ ($E_F= E_F^{(3)}$). These results highlight the robustness of our scheme against variations of the atomic filling factor.

\section{\label{sec:dynamics} Appendix B: Bloch oscillations and dynamics at arbitrary times: flat bands vs dispersive bands}

In this Appendix, we discuss the dynamics of the cloud at arbitrary times, so as to further reveal the interplay between the Hall drift taking place perpendicularly to the force $\bs E=E_y \bs 1_y$ -- due to the Berry's velocity $v_{\mathcal{F}}^x (\bs k)$ -- and the Bloch oscillations stemming from the band velocity $\bs v_{\text{band}}=(1/\hbar) \partial_{\bs k} E (\bs k)$, see main text.\\

\begin{figure}[h!]
\begin{center}
\includegraphics{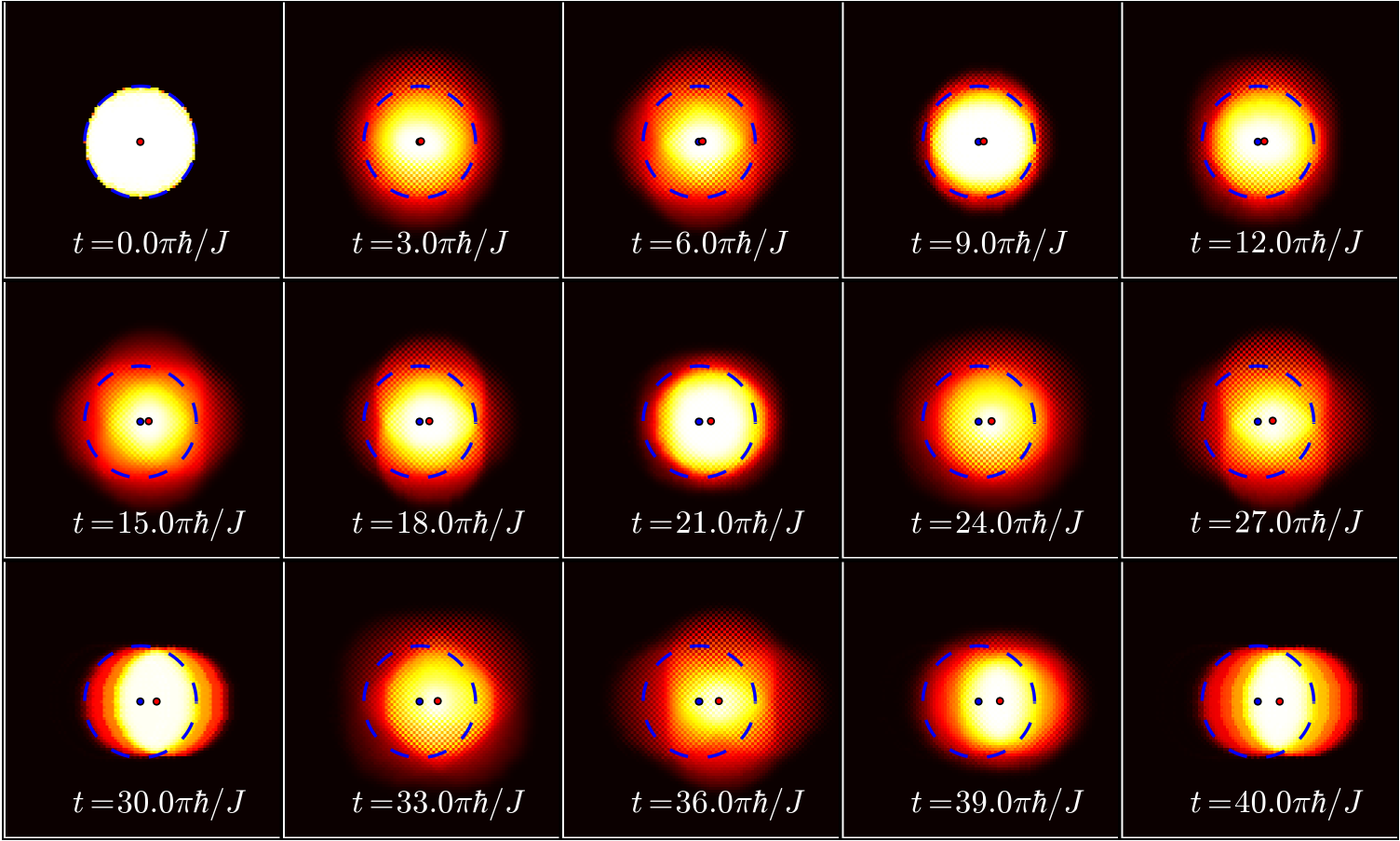}
\caption{Time-evolving density $\rho (\bs x,t)$ for a cloud initially trapped by an infinitely abrupt potential  $V_{\text{conf}} (r) = (r/r_0)^{\infty}$ with $r_0=20a$. The system parameters are $J_2= - 0.3J,\lambda_{\text{stag}}=0$, so that the system is initially prepared in a QH phase associated with the filled (dispersive) band $E_{+} (\bs k)$. The blue dashed circle and blue dot highlight the initial condition. The force applied after releasing the cloud is $\bs E=(0,0.2 J/a)$. }
\label{fig:secd_fig1}
\end{center}
\end{figure}

We first consider the case where the filled energy band is dispersive, in which case the contribution from the band velocity is large. To study such a situation, we start with the band structure depicted in Fig. \ref{fig:seca_fig1}(a) and reverse the sign of the hopping amplitude $J_2=0.3 J \rightarrow J_2= - 0.3 J$ so as to interchange (and reverse) the upper and lower bands $ E_{+} (\bs k) \leftrightarrow E_{-} (\bs k)$. Setting the Fermi energy in the gap, we now fill the dispersive band $E_{+} (\bs k)$, instead of the nearly flat band $E_{-} (\bs k)$. Note that the gap size $\Delta = 2J$ is the same as for the situation encountered in the main text ($J_2=0.3 J$). The time-evolving density is shown in Fig. \ref{fig:secd_fig1}, where large Bloch oscillations are observed between the periods $t= \text{integer} \times T$, where $T=2 \pi \hbar/a E_y=10 \pi \hbar/J$ is the time after which a full cycle is performed in the Brillouin zone (see main text). Note that these Bloch oscillations take place along both spatial directions, leading to a large broadening of the cloud at arbitrary times (see for example $t=6 \pi \hbar/J$ in Fig. \ref{fig:secd_fig1}). At $t= \text{integer} \times T$, the contribution of the band velocity vanishes, and the Hall drift is clearly visualized (see for example $t=40 \pi \hbar/J$ in Fig. \ref{fig:secd_fig1}). Note that the band $E_{+} (\bs k)$ is associated with the Chern number $\nu^{(+)}=-1$ (in contrast with $\nu^{(-)}=+1$ for $E_{-} (\bs k)$), which leads to a transverse displacement towards the right. The dispersive motion of the cloud can be analyzed through a semi-classical treatment, as already discussed in the main text.\\

We emphasize that, in general, the topological bulk bands produced in cold-atom systems will be dispersive. Consequently, the behavior presented in Fig. \ref{fig:secd_fig1}, showing a center-of-mass motion accompanied with Bloch oscillations, should correspond to the typical dynamics that will be observed in such experiments. \\

\begin{figure}[h!]
\begin{center}
\includegraphics{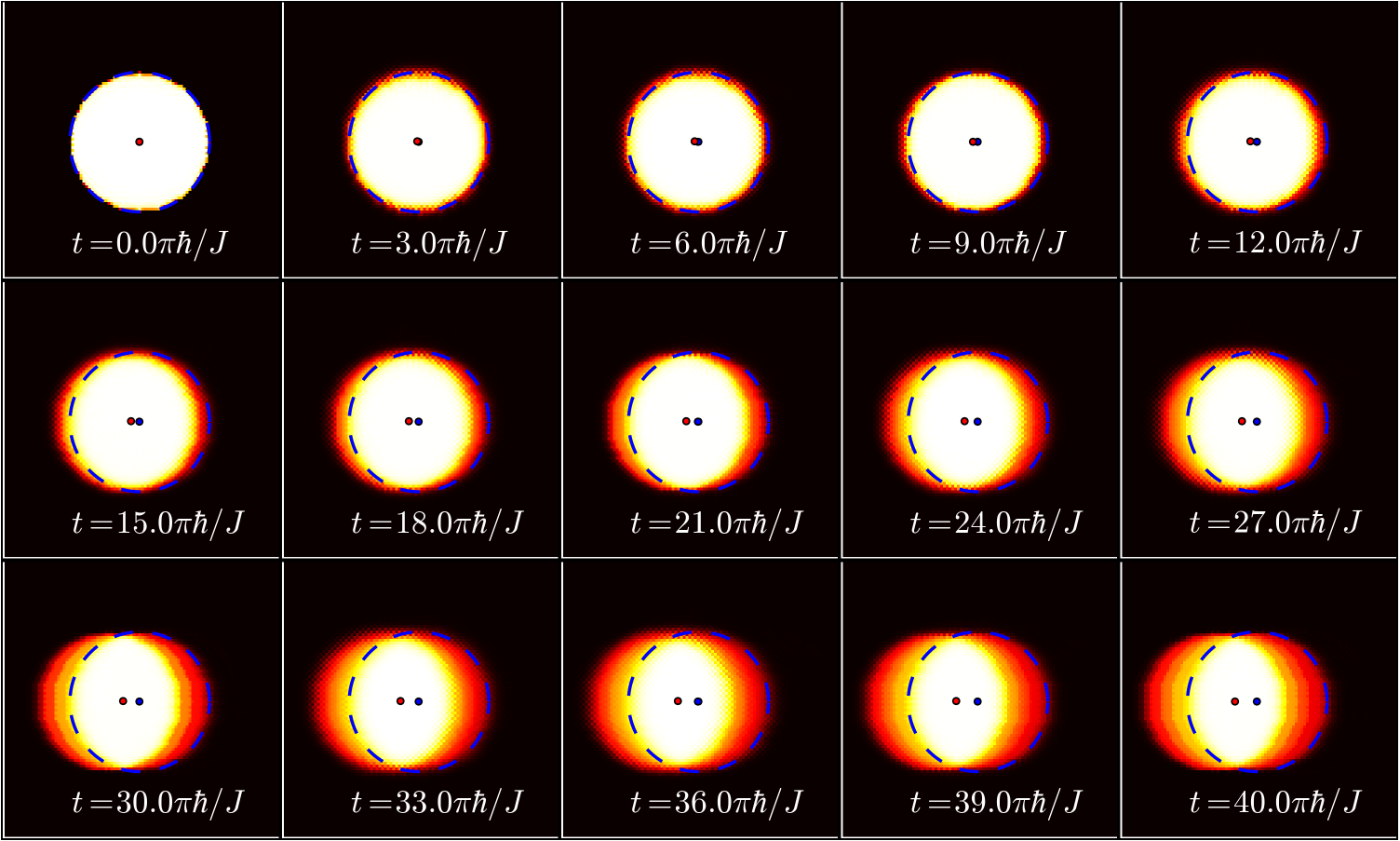}
\caption{Time-evolving density $\rho (\bs x,t)$ for a cloud initially trapped by an infinitely abrupt potential  $V_{\text{conf}} (r) = (r/r_0)^{\infty}$ with $r_0=25a$. The system parameters are $J_2= + 0.3J,\lambda_{\text{stag}}=0$, so that the system is initially prepared in a QH phase associated with the filled (nearly flat) band $E_{-} (\bs k)$. The blue dashed circle and blue dot highlight the initial condition. The force applied after releasing the cloud is $\bs E=(0,0.2 J/a)$. }
\label{fig:secd_fig2}
\end{center}
\end{figure}

To be complete, we show in Fig. \ref{fig:secd_fig2} the full dynamics in the case of the nearly flat band configuration obtained by setting $J_2=0.3 J$. In this case, the band velocity associated with the filled band $E_{-} (\bs k)$ is small, and thus, the Bloch oscillations only take place on the scale of a few lattice sites: the flat-band configuration reveals the Hall drift in a clear manner at arbitrary times.

\section{\label{sec:smoothpot}  Appendix C: Competition between trivial and non-trivial topological phases}

In the main text, we have shown that the Chern-number measurement allows to distinguish between trivial and non-trivial topological phases. These two different phases were obtained by either activating a staggered potential ($\lambda_\text{stag} \ne 0$ , $J_2=0$) or by activating the NNN-hopping term ($J_2 \ne 0$, $\lambda_\text{stag} = 0$), respectively.  However, it is instructive to study the case where both competing effects $J_2 , \lambda_\text{stag}$ are present, which can potentially give rise to either a trivial or a non-trivial topological phase. We have verified that our method still allows to precisely measure the Chern number $\nu_{\text{approx}} \simeq 0,\pm1$ in this situation, hence revealing the topological order of the atomic system. When $J_2=0.3 J$ and $\lambda_\text{stag}=0.3 J$, the system is in a topological phase characterized by a gap width $\Delta= 1.9 J$ and a Chern number $\nu=1$. The Chern number evaluated from the displacement $\bs x(t=40 \pi \hbar/J)$, using a force $E_y=0.2 J/a$, has been found to be $\nu_{\text{approx}} =1.00$ with less than $1\%$ error. Besides, when $J_2=0.3 J$ and $\lambda_\text{stag}=2.3 J$, the system is in an insulating state with the same gap width $\Delta = 1.9 J$ but with a vanishing Chern number; the measured $\nu_{\text{approx}} =0.00$ has been found with the same precision.


\section{\label{sec:smoothpot}  Appendix D: The Chern number measurement using smooth confinements}

\begin{figure}[h!]
\begin{center}
\includegraphics{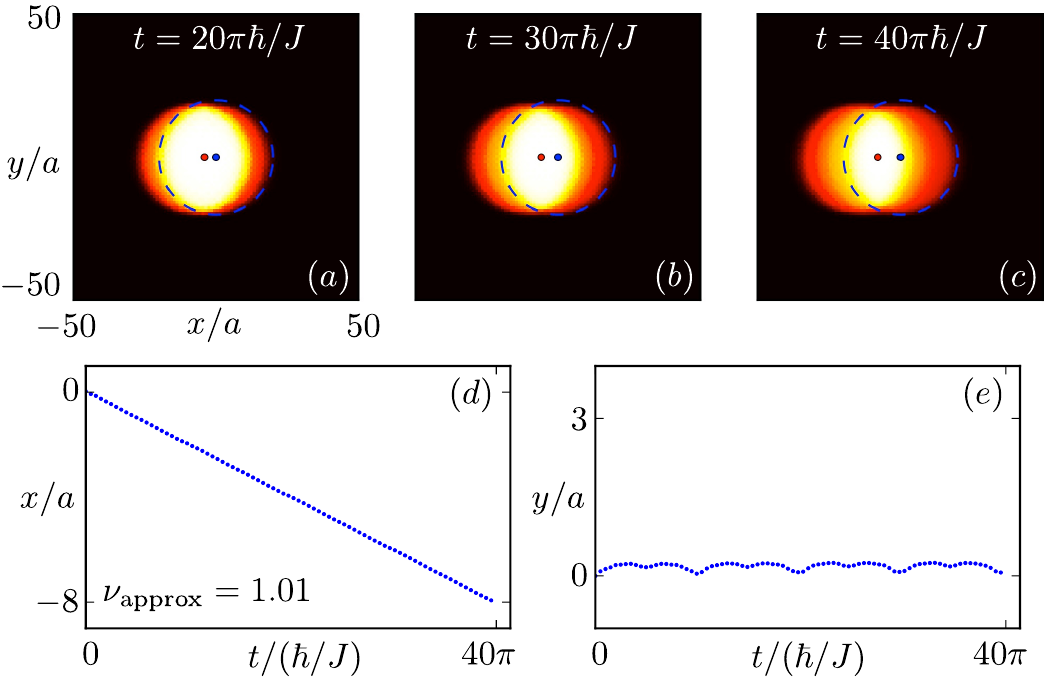}
\caption{Time-evolving density $\rho (\bs x,t)$ for a cloud initially trapped by a sharp potential  $V_{\text{conf}} (r) = 0.8 J (r/r_0)^{10}$ with $r_0=25a$. The system parameters are $J_2=0.3J,\lambda_{\text{stag}}=0$, so that the system is initially prepared in a QH phase. The blue dashed circle and blue dot highlight the initial condition. The force applied after releasing the cloud is $\bs E=(0,0.2 J/a)$. Figures (d)-(e) show the time evolution of the center of mass $\bs x (t)$. The Chern number deduced from Fig. (d) is indicated.}
\label{fig:secb_fig1}
\end{center}
\end{figure}

\begin{figure}[h!]
\begin{center}
\includegraphics{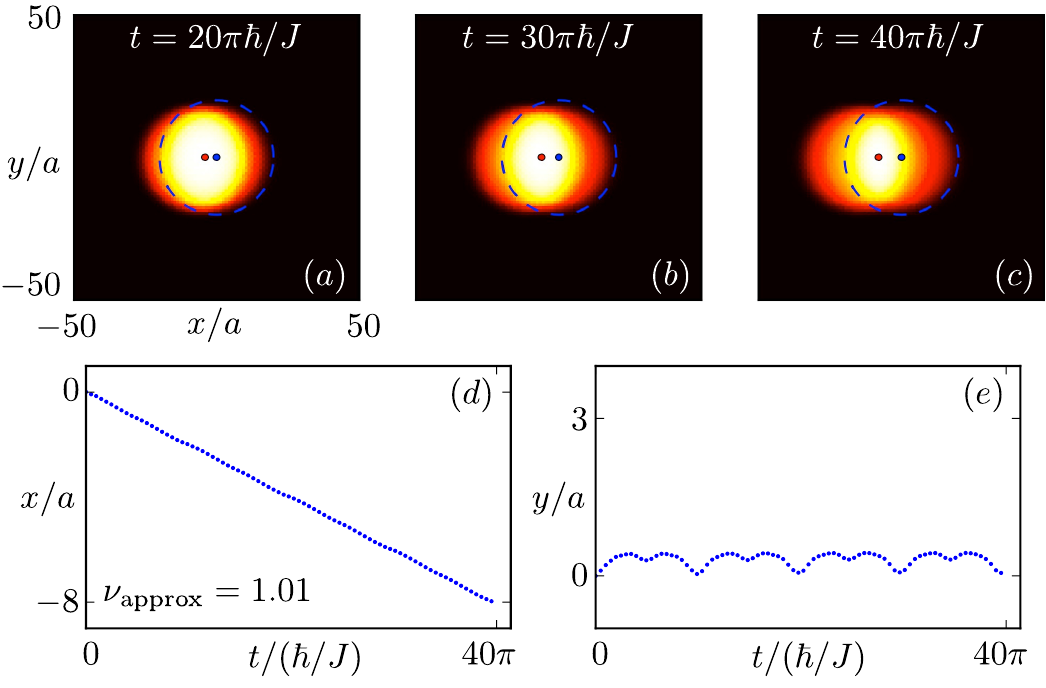}
\caption{Time-evolving density $\rho (\bs x,t)$ for a cloud initially trapped by a quartic potential  $V_{\text{conf}} (r) = 0.8 J (r/r_0)^{4}$ with $r_0=25a$. The system parameters are $J_2=0.3J,\lambda_{\text{stag}}=0$, so that the system is initially prepared in a QH phase. The blue dashed circle and blue dot highlight the initial condition. The force applied after releasing the cloud is $\bs E=(0,0.2 J/a)$. Figures (d)-(e) show the time evolution of the center of mass $\bs x (t)$. The Chern number deduced from Fig. (d) is indicated.}
\label{fig:secb_fig2}
\end{center}
\end{figure}

\begin{figure}[h!]
\begin{center}
\includegraphics{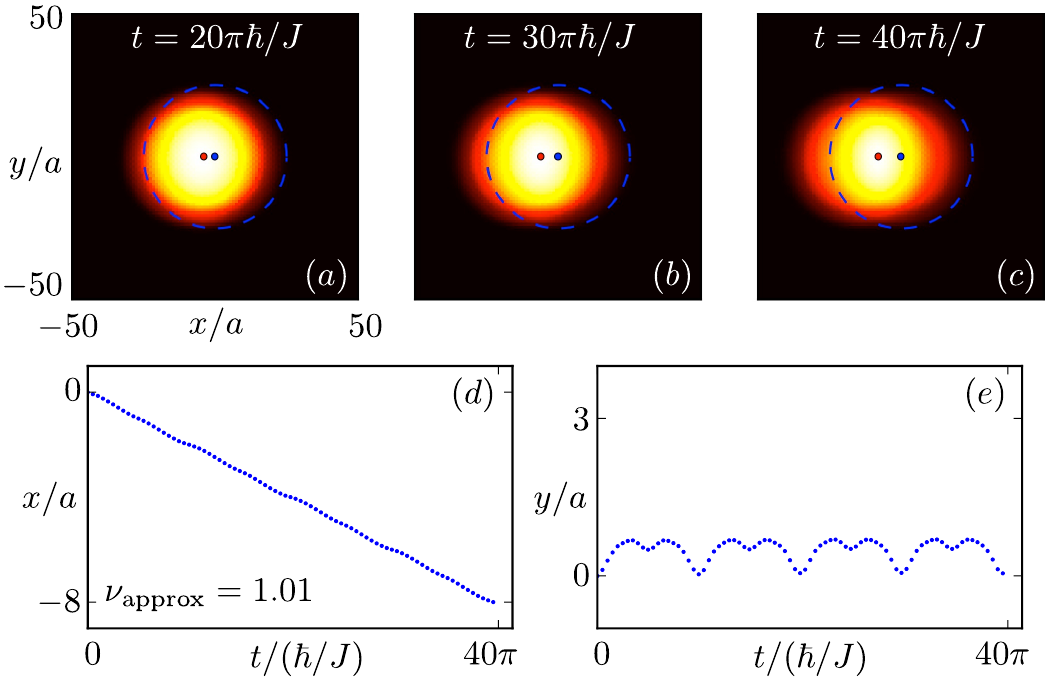}
\caption{Time-evolving density $\rho (\bs x,t)$ for a cloud initially trapped by a harmonic potential  $V_{\text{conf}} (r) = 0.8 J (r/r_0)^{2}$ with $r_0=25a$. The system parameters are $J_2=0.3J,\lambda_{\text{stag}}=0$, so that the system is initially prepared in a QH phase. The blue dashed circle and blue dot highlight the initial condition. The force applied after releasing the cloud is $\bs E=(0,0.2 J/a)$. Figures (d)-(e) show the time evolution of the center of mass $\bs x (t)$. The Chern number deduced from Fig. (d) is indicated.}
\label{fig:secb_fig3}
\end{center}
\end{figure}

\begin{figure}[h!]
\begin{center}
\includegraphics{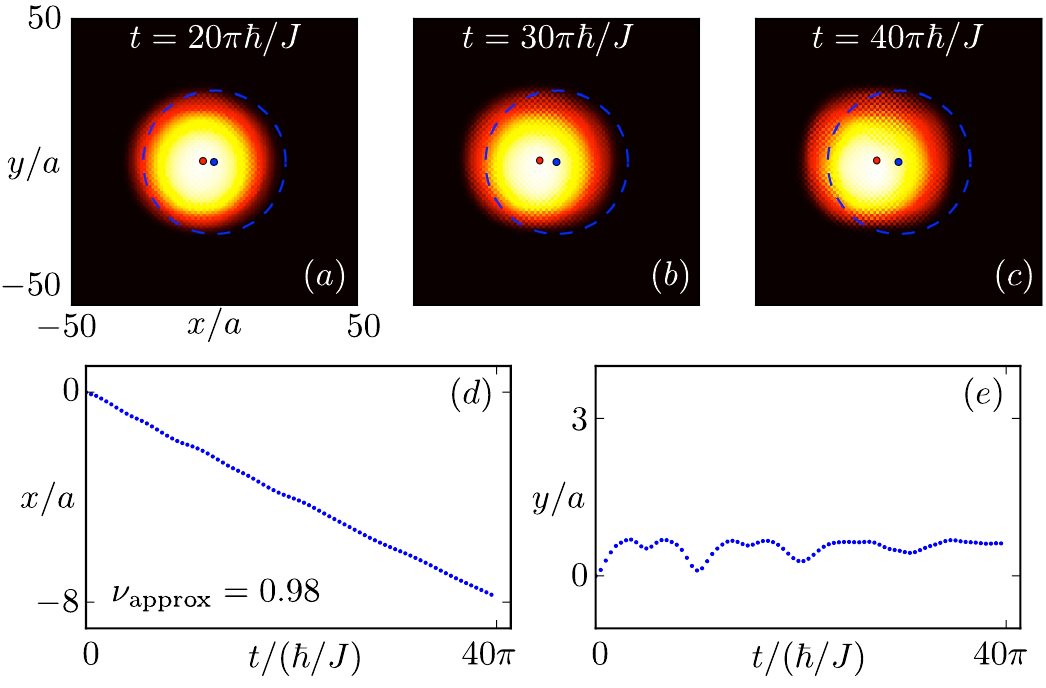}
\caption{Time-evolving density $\rho (\bs x,t)$ for a cloud initially trapped by a harmonic potential  $V_{\text{conf}} (r) = 0.8 J (r/r_0)^{2}$ with $r_0=25a$. At time $t=0$, the system is released in a weaker harmonic potential $V_{\text{conf}} (r) = 0.8 J (r/R_0)^{2}$, with $R_0=50a$. The system parameters are $J_2=0.3J,\lambda_{\text{stag}}=0$, so that the system is initially prepared in a QH phase. The blue dashed circle and blue dot highlight the initial condition. The force applied after releasing the cloud is $\bs E=(0,0.2 J/a)$. Figures (d)-(e) show the time evolution of the center of mass $\bs x (t)$. The Chern number deduced from Fig. (d) is indicated.}
\label{fig:secb_fig4}
\end{center}
\end{figure}

In the main text, we considered that the atomic cloud was initially trapped by an infinitely abrupt circular confinement, which was then suddenly removed at time $t=0$ when the force $\bs E = E_y \bs 1_y$ was applied. In this Appendix, we now study the time-evolved density $\rho (\bs x,t)$ in the situation where the initial confinement is chosen to be smooth, which is generally the case in most experiments. We performed numerical simulations for the following cases (setting the Fermi energy at the value $E_F=E_F^{(1)}=-1.3J$):
\begin{itemize}
\item A system initially confined by an abrupt potential $V_{\text{conf}} (r) = 0.8 J (r/r_0)^{10}$, see Fig. \ref{fig:secb_fig1};
\item A system initially confined by a quartic potential $V_{\text{conf}} (r) = 0.8 J (r/r_0)^{4}$, see Fig. \ref{fig:secb_fig2};
\item A system initially confined by a harmonic potential $V_{\text{conf}} (r) = 0.8 J (r/r_0)^{2}$, see Fig. \ref{fig:secb_fig3};
\item A system initially confined by a harmonic potential $V_{\text{conf}} (r) = 0.8 J (r/r_0)^{2}$, which is then suddenly released in a larger harmonic potential $V_{\text{conf}} (r) = 0.8 J (r/R_0)^{2}$ with $R_0 \gg r_0$, see Fig. \ref{fig:secb_fig4};
\end{itemize}
In all these situations, the atomic cloud shows a clear Hall drift along the $x$ direction, while the center of mass remains nearly immobile along the driven direction, $\vert y (t) \vert < a$. The Chern numbers deduced from Eq. \eqref{chern_position} remain remarkably close to the quantized value $\nu_{\text{approx}} \simeq \nu=1$, as indicated in all Figs. \ref{fig:secb_fig1}-\ref{fig:secb_fig4}. These numerical investigations demonstrate the applicability and robustness of our method in various confinement schemes.


\newpage

\begin{figure}[h!]
\begin{center}
\includegraphics{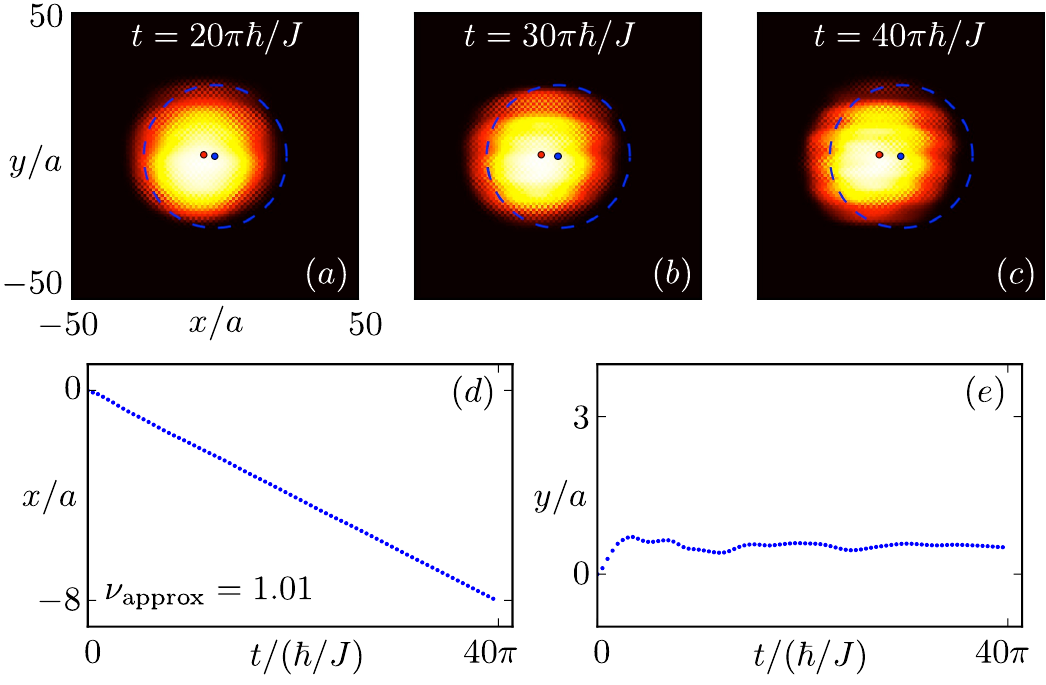}
\caption{Time-evolving density $\rho (\bs x,t)$ for a cloud initially trapped by a harmonic potential  $V_{\text{conf}} (r) = 0.8 J (r/r_0)^{2}$ with $r_0=25a$. At time $t=0$, the system is released along the  direction $x$ (transverse to the force $\bs E$), while it is still trapped along the $y$ direction by an anisotropic harmonic potential $V_{\text{conf}}^{(t)} (r) = 0.8 J (y/r_0)^{2}$. The system parameters are $J_2=0.3J,\lambda_{\text{stag}}=0$, so that the system is initially prepared in a QH phase. The blue dashed circle and blue dot highlight the initial condition. The force applied after releasing the cloud is $\bs E=(0,0.2 J/a)$. Figures (d)-(e) show the time evolution of the center of mass $\bs x (t)$. The Chern number deduced from Fig. (d) is indicated.}
\label{fig:secc_fig2}
\end{center}
\end{figure}

\begin{figure}[h!]
\begin{center}
\includegraphics{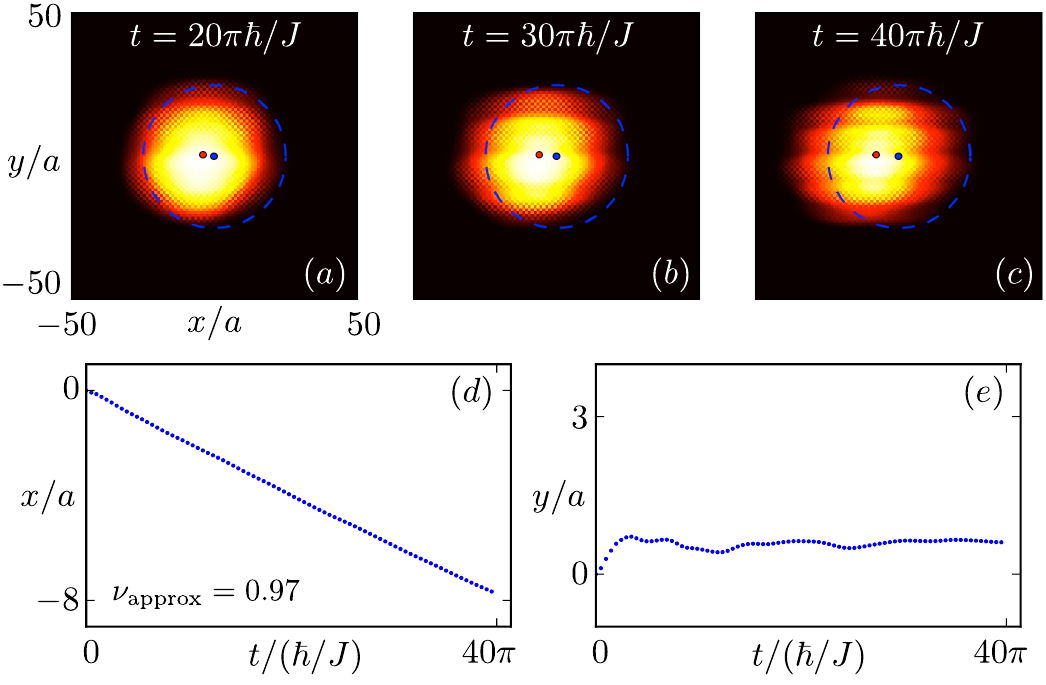}
\caption{Time-evolving density $\rho (\bs x,t)$ for a cloud initially trapped by a harmonic potential  $V_{\text{conf}} (r) = 0.8 J (r/r_0)^{2}$ with $r_0=25a$. At time $t=0$, the system is partially released along the  direction $x$ (transverse to the force $\bs E$), meaning that the cloud is suddenly confined by an anisotropic harmonic potential $V_{\text{conf}}^{(t)} (r) = 0.8 J \left [(x/R_0)^{2} + (y/r_0)^{2} \right ]$ with $R_0=50a$. The system parameters are $J_2=0.3J,\lambda_{\text{stag}}=0$, so that the system is initially prepared in a QH phase. The blue dashed circle and blue dot highlight the initial condition. The force applied after releasing the cloud is $\bs E=(0,0.2 J/a)$. Figures (d)-(e) show the time evolution of the center of mass $\bs x (t)$. The Chern number deduced from Fig. (d) is indicated.}
\label{fig:secc_fig1}
\end{center}
\end{figure}

\begin{minipage}{0.8\textwidth}

\section{\label{sec:releaseonedir} Appendix E: Releasing the confinement along the transverse direction only}

The topological order associated with the QH phase is captured by the Chern number $\nu$, which was shown to be deduced from the transverse motion of the center of mass $x(t)$ (the force being applied along the $y$ direction). Since no relevant information is contained in the longitudinal displacement $y(t)$, which might potentially perform Bloch oscillations, we may simplify the measurement scheme by simply releasing the cloud along the $x$ direction only. This possibility is investigated in this Appendix for two situations:

\begin{itemize}
\item  A system initially confined by a harmonic potential $V_{\text{conf}} (r) = 0.8 J (r/r_0)^{2}$ and released along the $x$ direction only: at time $t=0$ the cloud is confined by the anisotropic potential $V_{\text{conf}}^{(t)} (r) = 0.8 J (y/r_0)^{2}$, while the force is applied along the $y$ direction; see Fig. \ref{fig:secc_fig2};
\item  A system initially confined by a harmonic potential $V_{\text{conf}} (r) = 0.8 J (r/r_0)^{2}$ and \emph{only partially} released along the $x$ direction: at time $t=0$ the cloud is confined by the anisotropic potential $V_{\text{conf}}^{(t)} (r) = 0.8 J \left [(x/R_0)^{2} + (y/r_0)^{2} \right ]$ with $R_0 > r_0$, while the force is applied along the $y$ direction; see Fig. \ref{fig:secc_fig1};
\end{itemize}

The Chern number deduced from these two anisotropic schemes remains close to the quantized value $\nu_{\text{approx}} \simeq \nu=1$, indicating the validity of our method in these situations.

\end{minipage}

\end{document}